\documentclass[3p,times]{elsarticle}
\usepackage{graphicx}
\usepackage{amssymb} 
\usepackage{verbatim}
\usepackage{amsthm} 
\usepackage{lineno}
\usepackage{harpoon}

\usepackage{hyperref}
\hypersetup{linkcolor=blue,citecolor=blue,filecolor=black,urlcolor=blue}
\newcommand{\pT} {\ensuremath{p_{\mathrm{T}}}}

\journal{Nuclear Physics A} 

\begin{document}

\begin{frontmatter} 

% Your Title - please insert
\title{Measurement of the distributions of event-by-event flow harmonics in Pb-Pb Collisions at $\sqrt{s_{\mathrm{NN}}}$=2.76 TeV with the ATLAS detector}

%% Single author (and collaboration) - please insert
\author{Jiangyong Jia (on behalf of the ATLAS\fnref{col1} Collaboration)}
\fntext[col1] {A list of members of the ATLAS Collaboration and acknowledgements can be found at the end of this issue.}
\address{Chemistry Department, Stony Brook University, NY 11794, USA; Physics Department, Brookhaven National Laboratory, NY 11796, USA}

%% Multiple authors
%\author[auth2]{Marcus Junius Brutus}
%\address[auth1]{Somewhere, Rome}
%\address[auth2]{Somewhere else, Rome}
\begin{abstract} 
The event-by-event distributions of harmonic flow coefficients $v_n$ for $n$=2--4 are measured in Pb-Pb collisions at $\sqrt{s_{NN}}=2.76$ TeV  with the ATLAS detector, using charged particles with $\pT> 0.5$ GeV and $|\eta|<2.5$. The shape of the $v_n$ distributions is consistent with Gaussian fluctuations in central collisions for $v_2$ and over the measured centrality range for $v_3$ and $v_4$. When these distributions are rescaled to have the same $\langle v_n\rangle$, the resulting shapes are nearly the same for $\pT>1$ GeV and $0.5<\pT<1$ GeV. The $v_n$ distributions are compared with the eccentricity distributions from two initial geometry models: Glauber and the MC-KLN model. Both fail to describe the $v_2$ data consistently over the full centrality range.
\end{abstract} 

\end{frontmatter} % do not change

%% linenumbers are useful for reviewing process
%\linenumbers

In recent years, the measurement of harmonic flow coefficients $v_n$ has provided important insight into the hot and dense matter created in heavy ion collisions at the Relativistic Heavy Ion Collider (RHIC) and the Large Hadron Collider (LHC). These coefficients are generally obtained from a Fourier expansion of particle azimuthal angle distributions, $\frac{dN}{d\phi}\propto1+2\sum_{n=1}^{\infty}v_{n}\cos n(\phi-\Phi_{n})$, where $\Phi_n$ represents the phase of $v_n$ (event plane or EP)~\cite{Aad:2012bu,Jia:2012sa}. Previous measurements determined the $v_n$ from the distribution of $\phi-\Phi_{n}$, accumulated over many events. This event-averaged $v_n$ mainly reflects the hydrodynamic response of the created matter to the average collision geometry in the initial state. However, more information can be obtained by measuring the $v_n$ on a event-by-event (EbE) basis. This is particularly useful for understanding the nature of the EbE fluctuations in the initial collision geometry. This proceedings present the methods and results of EbE $v_n$ for $n=2-4$ obtained with the ATLAS detector~\cite{Aad:2008zzm}. 

\begin{figure}[h]
\vspace*{-0.2cm}\begin{tabular}{lr}
\begin{minipage}{0.77\linewidth}
\begin{flushleft}
\includegraphics[width=1\columnwidth]{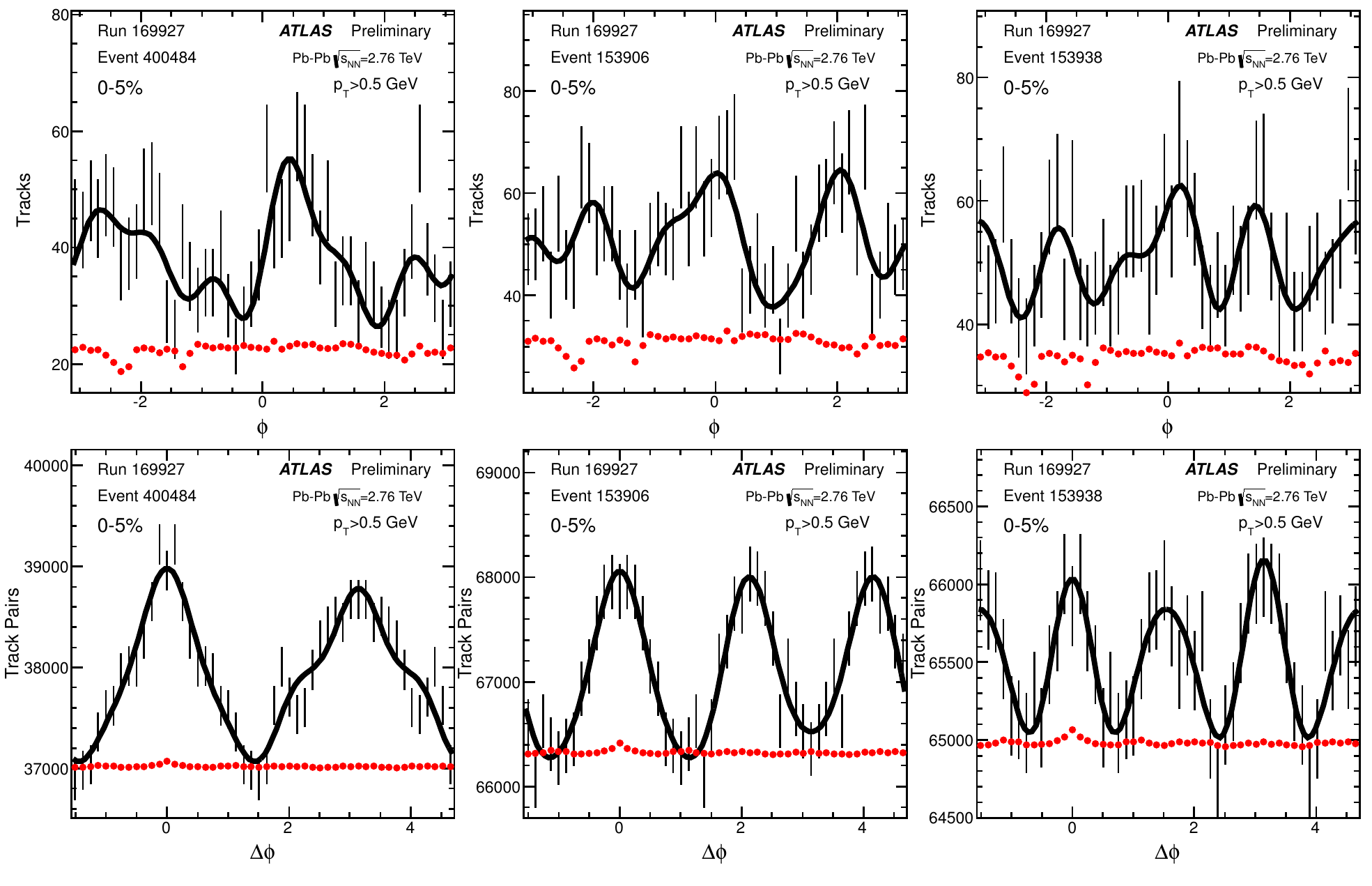}\vspace*{-0.6cm}
\end{flushleft}
\end{minipage}
\hspace*{0.2cm}\begin{minipage}{0.19\linewidth}
\begin{flushright}
\caption{\label{fig:1} Single track $\phi$ (top) and track pair $\Delta\phi$ (bottom) distributions for three events (from left to right) in 0-5\% centrality interval~\cite{ebe}. The pair distributions have been folded into [-0.5$\pi$,1.5$\pi$]. The bars indicate the statistical uncertainties, the solid curves indicate a Fourier parameterization including first six harmonics: $dN/d\phi=A(1+2\sum_{i=1}^{6}c_n\cos n(\phi-\Psi_n))$ for single track distributions and $dN/d\Delta\phi=A(1+2\sum_{i=1}^{6}c_n\cos n(\Delta\phi))$ for pair distributions, and the red solid points indicate the acceptance functions (arbitrary normalization).}
\end{flushright}
\end{minipage}
\end{tabular}
\vspace*{-0.2cm}\end{figure}

This analysis is based on 8~$\mu\mathrm{b}^{-1}$ of minimum bias Pb-Pb data collected in 2010 at $\sqrt{s_{_{\mathrm{NN}}}}=2.76$~TeV~\cite{ebe}. The $v_n$ coefficients are calculated using tracks with $\pT>0.5$ GeV and $|\eta|<2.5$, reconstructed by the inner detector (ID). To illustrate the level of EbE fluctuations, Figure~\ref{fig:1} shows the distributions of track $\phi$ and track pair relative angle $\Delta\phi$ with $\pT>0.5$ GeV for three central events. Rich EbE patterns, beyond the structures of the detector acceptance (solid points) and statistical fluctutions, are observed. These distributions are the inputs for the EbE $v_n$ analyses.

Two methods are used to obtain the EbE $v_n$. The first method starts with a Fourier expansion of the azimuthal distribution of charged particles in a given event, reweighted by the inverse of the tracking efficiency for each particle:
\small{
\begin{eqnarray}
\label{eq:1}
\frac{dN}{d\phi}\propto 1+2\sum_{n=1}^{\infty}v_n^{\mathrm{obs}}\cos n(\phi-\Phi_{n}^{\mathrm{obs}})=  1+2\sum_{n=1}^{\infty}\left(v_{n,\mathrm{x}}^{\mathrm{obs}}\cos n\phi+v_{n,\mathrm{y}}^{\mathrm{obs}}\sin n\phi)\right), v_n^{\mathrm{obs}} = \sqrt{\left(v_{n,\mathrm{x}}^{\mathrm{obs}}\right)^2+\left(v_{n,\mathrm{y}}^{\mathrm{obs}}\right)^2},\mathrm{tan} n\Phi_{n}^{\mathrm{obs}}=\frac{v_{n,\mathrm{y}}^{\mathrm{obs}}}{v_{n,\mathrm{x}}^{\mathrm{obs}}}
\end{eqnarray}}\normalsize
where $v_n^{\mathrm{obs}}$ is the magnitude of the observed per-particle flow vector: $\overrightharp{v}_n^{\;\mathrm{obs}}=(v_{n,\mathrm{x}}^{\mathrm{obs}},v_{n,\mathrm{y}}^{\mathrm{obs}})$. In the limit of infinite multiplicity and the absence of non-flow effects, it approaches the true flow signal: $v_n^{\mathrm{obs}}\rightarrow v_n$. The key of the measurement is to determine the response function $p(\overrightharp{v}_n^{\;\mathrm{obs}}|\overrightharp{v}_n)$ or $p(v_n^{\mathrm{obs}}|v_n)$, used to unfold these smearing effects.

In order to determine the response function, the tracks in the ID are divided into two subevents with symmetric $\eta$ range, $\eta>0$ and $\eta<0$. The smearing effects are estimated from the difference of the flow vectors between the two subevents, $p_{\mathrm{sub}}\left((\overrightharp{v}_n^{\;\mathrm{obs}})^a-(\overrightharp{v}_n^{\;\mathrm{obs}})^b\right)$, for which the physical flow signal cancels. This distribution is observed to be well described by a 2-D Gaussian with identical widths in both dimensions, $\delta_{_{\mathrm{2SE}}}$, and hence the 2-D response function can be obtained as $p(\overrightharp{v}_n^{\;\mathrm{obs}}|\overrightharp{v}_n)=cp_{\mathrm{sub}}\left(c\left[(\overrightharp{v}_n^{\;\mathrm{obs}})^a-(\overrightharp{v}_n^{\;\mathrm{obs}})^b\right]\right)$, with $c=2$ and $\sqrt{2}$ for the full-ID and half-ID, respectively. A simple shift of this 2-D distribution to $\overrightharp{v}_n=(v_{n,\mathrm x},v_{n,\mathrm y})$, followed by an integration over the azimuthal angle, gives the desired 1-D response function in the form of the well-known Bessel-Gaussian function~\cite{Ollitrault:1992bk}:
\begin{eqnarray}
\label{eq:5b}
p(v_n^{\mathrm{obs}}|v_n)\propto v_n^{\mathrm{obs}}e^{-\frac{(v_n^{\mathrm{obs}})^2+v_n^2}{2\delta^2}} I_0\left(\frac{v_n^{\mathrm{obs}}v_n}{\delta^2}\right),  \delta= &\left\{\begin{array}{ll}
  \delta_{_{\mathrm{2SE}}}/\sqrt{2}  &\textrm{ for half-ID } 
    \\
   \delta_{_{\mathrm{2SE}}}/2  &\textrm{ for full-ID } 
    \end{array}\right. \,
\end{eqnarray}
where $I_0$ is the modified Bessel function of the first kind and the difference $s=v_n^{\mathrm{obs}}-v_n$ accounts for the smearing. 

In the second method, the observed flow signal is defined from an EbE pair distribution, obtained by convolving the tracks in the first half-ID with those in the second half-ID:
\small{\begin{eqnarray}
%\label{eq:p1}
\nonumber
&&\hspace*{-0.7cm}\frac{dN}{d\Delta\phi} \propto \left[1+2\sum_n\left(v_{n,\mathrm{x}}^{\mathrm{obs_1}}\cos n\phi_1+ v_{n,\mathrm{y}}^{\mathrm{obs_1}}\sin n\phi_1\right)\right]\otimes\left[1+2\sum_n\left(v_{n,\mathrm{x}}^{\mathrm{obs_2}}\cos n\phi_2+ v_{n,\mathrm{y}}^{\mathrm{obs_2}}\sin n\phi_2\right)\right] = 1+2\sum_n\left(A_n\cos n\Delta\phi+ B_n\sin n\Delta\phi\right)\\
\label{eq:p2}
&&\hspace*{-0.5cm}v_n^{\mathrm{obs,2PC}} = \left(A_n^2+B_n^2\right)^{1/4} = \left[\left(v_{n,\mathrm{x}}^{\mathrm{obs_1}}\right)^2+\left(v_{n,\mathrm{y}}^{\mathrm{obs_1}}\right)^2\right]^{1/4}\left[\left(v_{n,\mathrm{x}}^{\mathrm{obs_2}}\right)^2+\left(v_{n,\mathrm{y}}^{\mathrm{obs_2}}\right)^2\right]^{1/4}= \sqrt{v_n^{\mathrm{obs_1}}v_n^{\mathrm{obs_2}}}= \sqrt{(v_n+s_1)(v_n+s_2)}\;
\end{eqnarray}}\normalsize
where $s_1=v_n^{\mathrm{obs_1}}-v_n$ and $s_2=v_n^{\mathrm{obs_2}}-v_n$ are independent variables described by the probability distribution in Eq.~\ref{eq:5b} with $\delta=\delta_{_{\mathrm{2SE}}}/\sqrt{2}$. The response function for $v_n^{\mathrm{obs,2PC}}$ is different from $v_n^{\mathrm{obs}}$ due to the presence of two random variables.

The Bayesian unfolding procedure from~\cite{Agostini} is used to calculate the $v_n$ distribution, in three $\pT$ ranges: $\pT>0.5$ GeV, $1>\pT>0.5$ GeV and $\pT>1$ GeV. In each case, the prior (initial distribution) is taken as the $v_{n}^{\mathrm{obs}}$ distribution from the full-ID, and the number of iterations $N_{\mathrm{iter}}$ is adjusted according to the sample statistics and binning. The convergence is generally reached for $N_{\mathrm{iter}}\ge8$ in the case of $n=2$, but more iterations are required for larger $n$ and in more peripheral collisions. The final results are found to be independent of the choice of priors, although the convergence is faster with an appropriate choice. The robustness of the unfolding is checked by comparing the results measured independently for the half-ID and the full-ID, as well as by comparing between the two unfolding methods. Despite the large differences in their initial distributions, the final unfolded results always agree. The final results are also found to be insensitive to the explicit requirement of a $\eta$ gap between the two subevents in both methods, suggesting that the influences of the short-range non-flow correlations on the final results are not significant.

\begin{figure}[h]
\begin{tabular}{lr}
\begin{minipage}{0.77\linewidth}
\begin{flushleft}
\includegraphics[width=1\columnwidth]{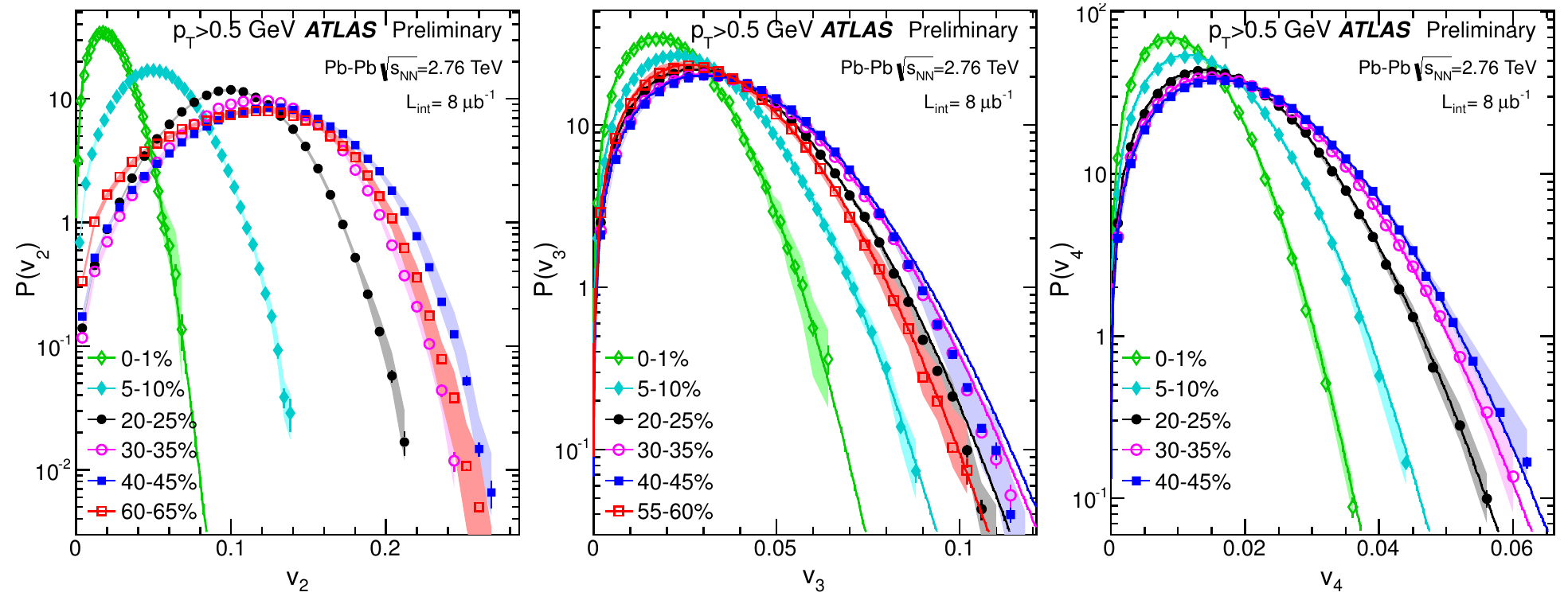}
\vspace*{-0.2cm}
\end{flushleft}
\end{minipage}
\hspace*{0.2cm}
\begin{minipage}{0.18\linewidth}
\begin{flushright}
\caption{\label{fig:result1} The probability distribution of the EbE $v_n$ for $n=2$ (left), $n=3$ (middle) and $n=4$ (right)~\cite{ebe}. The solid curves are Gaussian distributions with mean adjusted to the measured $\langle v_n\rangle$, shown for the 0-1\% centrality interval for $v_2$, but for all centrality intervals for $v_3$ and $v_4$.}
\end{flushright}
\end{minipage}
\end{tabular}
\vspace*{-0.4cm}
\end{figure}

Figure~\ref{fig:result1} shows the probability distribution of the EbE $v_n$ in several centrality intervals obtained for charged particles at $\pT>0.5$ GeV. The shape of these distributions changes strongly with centrality for $v_2$, while it is relatively unchanged for higher-order harmonics. These distributions are compared with the PDF obtained by radial projection of a 2-D Gaussian distribution in $\overrightharp{v}_n$: $P(v_n)=\frac{v_n}{\sigma^2}e^{-\frac{v_n^2}{2\sigma^2}}, \sigma=\sqrt{\frac{2}{\pi}}\langle v_n\rangle$. The Gaussian description works well for $v_3$ and $v_4$ over the measured centrality range, but fails for $v_2$ beyond the top 2\% most central collisions.

\begin{figure}[h]
\begin{tabular}{lr}
\begin{minipage}{0.61\linewidth}
\begin{flushleft}
\includegraphics[width=1\columnwidth]{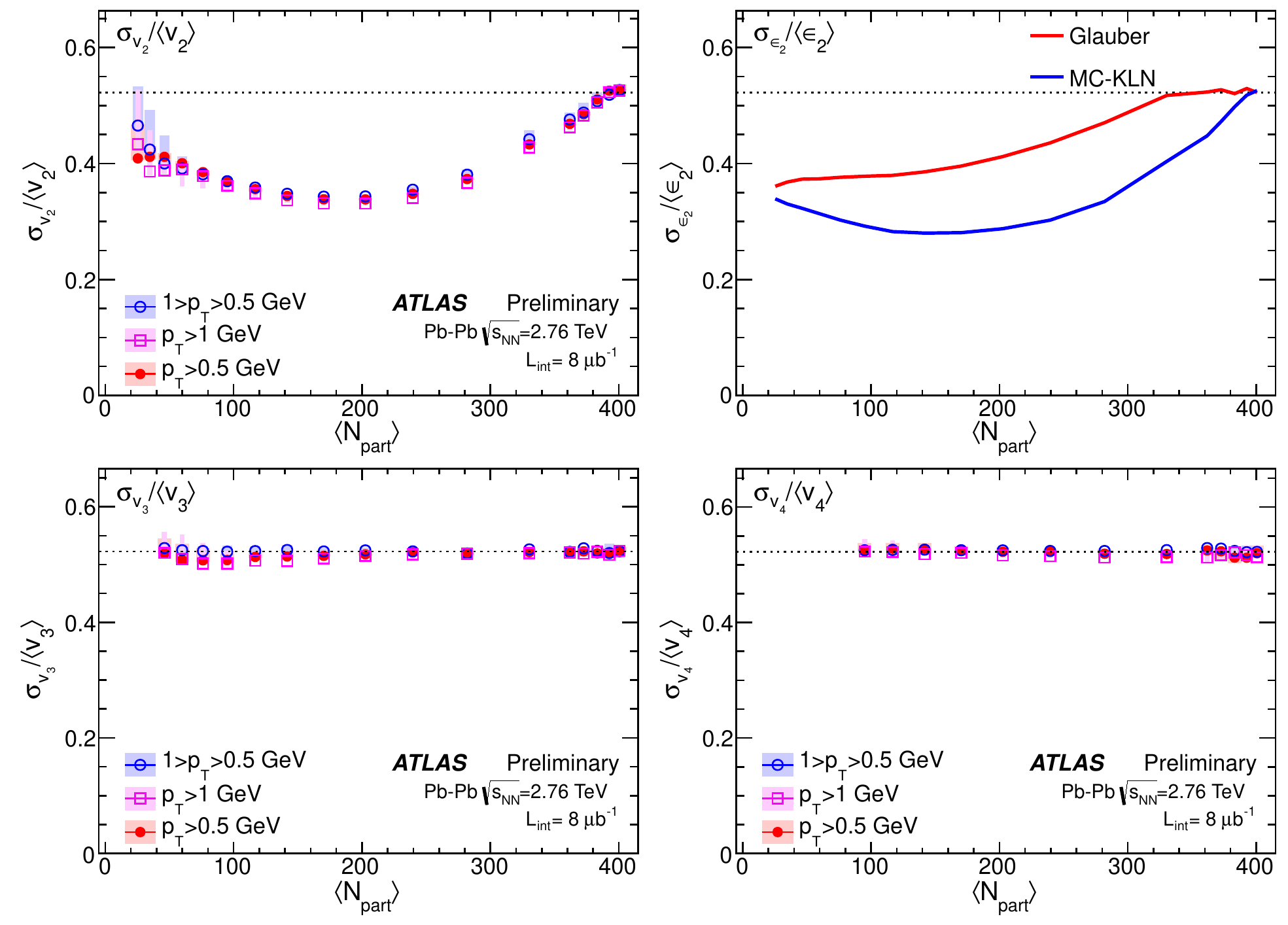}
\vspace*{-0.2cm}
\end{flushleft}
\end{minipage}
\hspace*{0.4cm}
\begin{minipage}{0.25\linewidth}
\begin{flushright}
\caption{\label{fig:result2}  The $\sigma_{v_n}/\langle v_n\rangle$ vs. $\langle N_{\mathrm{part}}\rangle$ in three $\pT$ ranges for $n=2$ (top-left), $n=3$ (bottom-left) and $n=4$ (bottom-right)~\cite{ebe}. The dotted lines indicate $\sqrt{4/\pi-1}\approx0.523$ expected for Gaussian fluctuations. Top-right panel shows the $\sigma_{\epsilon_2}/\langle\epsilon_2\rangle$ for the Glauber model~\cite{Miller:2007ri} and the MC-KLN model~\cite{Drescher:2006pi}.}
\end{flushright}
\end{minipage}
\end{tabular}
\vspace*{-0.4cm}
\end{figure}

Many quantities can be calculated directly from these distributions, such as the mean $\langle v_n\rangle$, width $\sigma_{v_n}$, ratio $\sigma_{v_n}/\langle v_n\rangle$ and RMS value $\sqrt{\langle v_n^2\rangle}\equiv\sqrt{\langle v_n\rangle^2+\sigma_{v_n}^2}$. The $\sigma_{v_n}/\langle v_n\rangle$ is a measure of the relative fluctuations of $v_n$ and was previously estimated indirectly from the two- and four-particle cumulant methods~\cite{Alice:2011yba}. Figure~\ref{fig:result2} shows the $\sigma_{v_n}/\langle v_n\rangle$ calculated for the three $\pT$ ranges. Despite the strong $\pT$ dependence of $\langle v_n\rangle$ and $\sigma_{v_n}$, the ratios are remarkably stable, suggesting that the hydrodynamic response to the initial geometry is nearly independent of $\pT$. For $v_2$, the values of $\sigma_{v_n}/\langle v_n\rangle$ vary strongly with $\langle N_{\mathrm{part}}\rangle$, and reach a minimum of about 0.34 at $\langle N_{\mathrm{part}}\rangle\sim200$ or 20-30\% centrality range. For $v_3$ and $v_4$, the values of $\sigma_{v_n}/\langle v_n\rangle$ are almost independent of $\langle N_{\mathrm{part}}\rangle$, and are consistent with the value expected from Gaussian distributions ($\sqrt{4/\pi-1}\approx0.523$ as indicated by the dotted lines). 

\begin{figure}[h]
\begin{tabular}{lr}
\begin{minipage}{0.72\linewidth}
\begin{flushleft}
 \includegraphics[width=1\columnwidth]{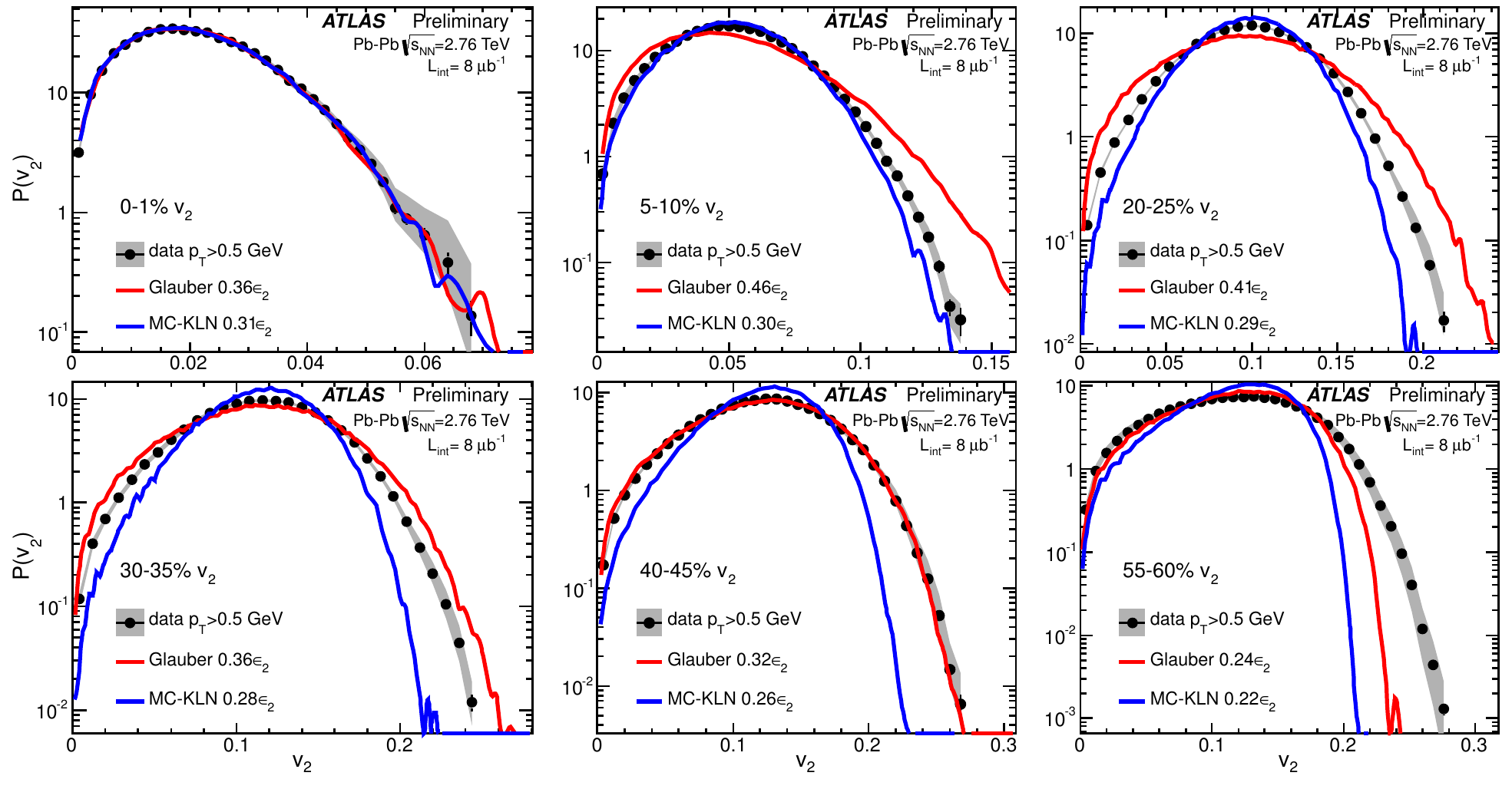}
\vspace*{-0.2cm}
\end{flushleft}
\end{minipage}
\hspace*{0.2cm}
\begin{minipage}{0.22\linewidth}
\begin{flushright}
\caption{\label{fig:result3} The EbE $v_2$ distributions compared with the $\epsilon_2$ distributions from the Glauber model (red lines) and the MC-KLN model (blue lines)~\cite{ebe}.}
\end{flushright}
\end{minipage}
\end{tabular}
\vspace*{-0.5cm}
\end{figure}

Figure~\ref{fig:result3} compares the EbE $v_2$ distributions with the distributions of the eccentricity $\epsilon_2$ of the initial geometry, calculated for the Glauber model~\cite{Miller:2007ri} and the MC-KLN model (version 3.46)~\cite{Drescher:2006pi}. The $\epsilon_2$ distribution for each centrality interval is rescaled to match the $\langle v_2\rangle$ of the data, and then normalized into a PDF. Figure~\ref{fig:result3} shows that the rescaled $\epsilon_2$ distributions describe the data well for the most central collisions, but start to fail in non-central collisions. This behavior is also reflected in the comparison of $\sigma_{v_2}/\langle v_2\rangle$ with $\sigma_{\epsilon_2}/\langle \epsilon_2\rangle$ in the top-right panel of Figure~\ref{fig:result2}. The agreement with the models for $n=3-4$ (see~\cite{ebe} for more details) are better than the $n=2$ case, however, this could simply reflect the fact that all distributions are dominated by Gaussian fluctuations, which have a universal shape.
% \begin{figure}[!h!]
% \centering
% \includegraphics[width=0.8\columnwidth]{cshape2GLrnprelim_har2_reb2_finalcomb}
% \caption{\label{fig:result3} The EbE $v_2$ distributions compared with the $\epsilon_2$ distributions from the Glauber model (red lines) and the MC-KLN model (blue lines).}
% \end{figure}

\begin{figure}[h]
\begin{tabular}{lr}
\begin{minipage}{0.72\linewidth}
\begin{flushleft}
\includegraphics[width=1\columnwidth]{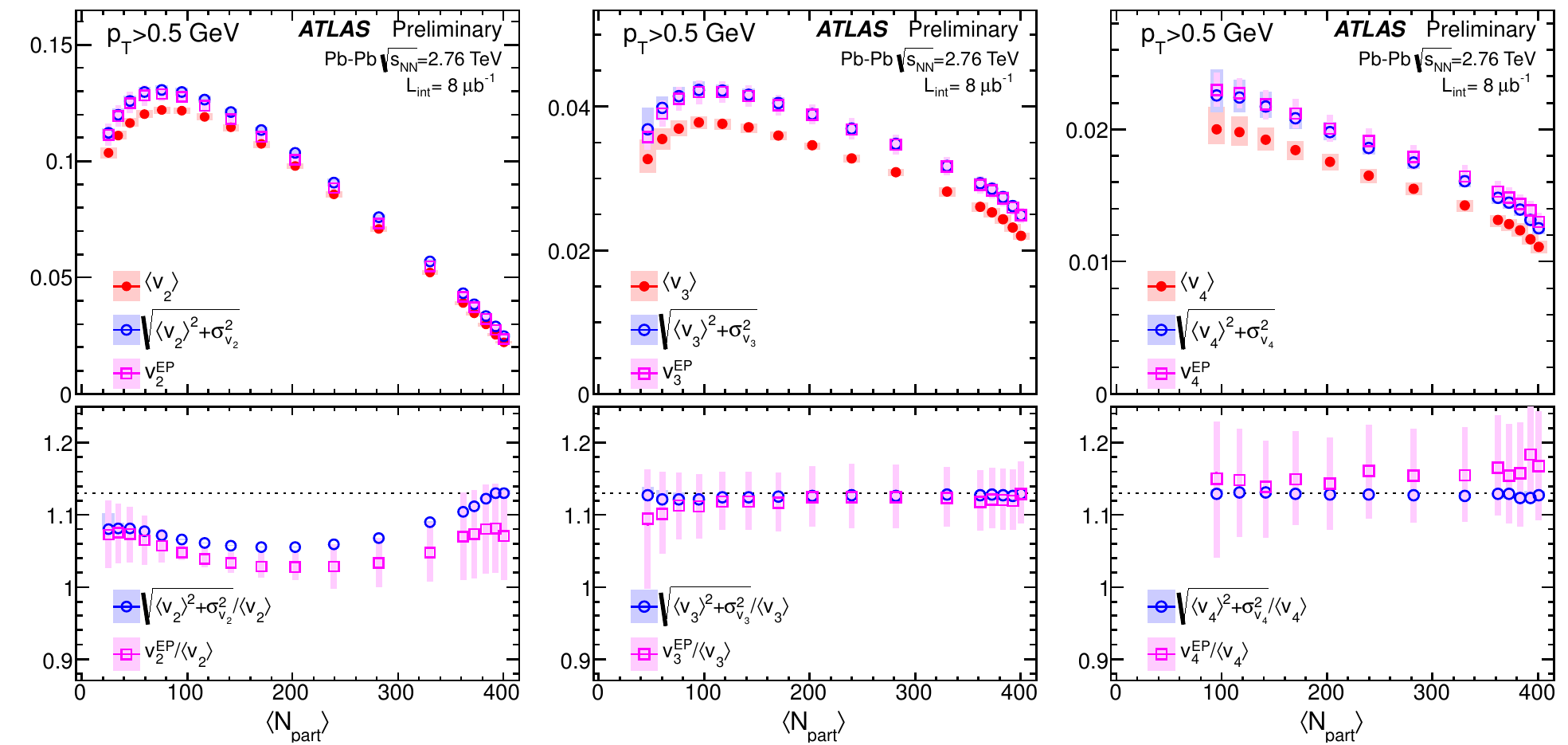}
\vspace*{-0.2cm}
\end{flushleft}
\end{minipage}
\hspace*{0.2cm}
\begin{minipage}{0.20\linewidth}
\begin{flushright}
 \caption{\label{fig:result4} Top panels: Comparison of $\langle v_n\rangle$ and $\sqrt{\langle v_n^2\rangle}\equiv\sqrt{\langle v_n\rangle^2+\sigma_{v_n}^2}$ with $v_n^{\mathrm{EP}}$~\cite{ebe}. Bottom panels: the ratios of $\sqrt{\langle v_n^2\rangle}$ and $v_n^{\mathrm{EP}}$ to $\langle v_n\rangle$~\cite{ebe}. The dotted lines in bottom panels at $\sqrt{\langle v_n^2\rangle}/\langle v_n\rangle=\sqrt{4/\pi}\approx1.13$ indicate the value expected for Gaussian fluctuations.}
\end{flushright}
\end{minipage}
\end{tabular}
\vspace*{-0.4cm}
\end{figure}

%The lower limit is reached when the resolution factor~\cite{Aad:2012bu} used in the EP method is one, while the upper limit is reached when the resolution factor is close to zero. 
The EP method in general is known to measure a $v_n$ value between the simple average and the RMS of the true $v_n$~\cite{Alver:2008zza}: $\langle v_n\rangle\leq v_n^{\mathrm{EP}}\leq\sqrt{\langle v_n^2\rangle}$. This relation is checked explicitly in Figure~\ref{fig:result4} based on the EbE $v_n$ distributions. For $v_3$ and $v_4$, the values of $v_n^{\mathrm{EP}}$ are almost identical to $\sqrt{\langle v_n^2\rangle}$; For $v_2$, the values of $v_2^{\mathrm{EP}}$ are in between $\langle v_n\rangle$ and $\sqrt{\langle v_n^2\rangle}$: they are closer to $\langle v_n\rangle$ in mid-central collisions where the EP resolution factor is close to one, and approach $\sqrt{\langle v_n^2\rangle}$ in peripheral collisions where the resolution factor is small.
% \begin{figure}[!h!]
% \centering
% \includegraphics[width=0.8\columnwidth]{vnscalsysfinalEPR_w1}
% \caption{\label{fig:result4} Top panels: Comparison of $\langle v_n\rangle$ and $\sqrt{\langle v_n^2\rangle}\equiv\sqrt{\langle v_n\rangle^2+\sigma_{v_n}^2}$ with $v_n^{\mathrm{EP}}$. Bottom panels: the ratios of $\sqrt{\langle v_n^2\rangle}$ and $v_n^{\mathrm{EP}}$ to $\langle v_n\rangle$. The dotted lines in bottom panels at $\sqrt{\langle v_n^2\rangle}/\langle v_n\rangle=1.13$ indicate the value expected for Gaussian fluctuations.}
% \end{figure}

In summary, the EbE $v_n$ distributions for $n=2-4$ are measured for Pb-Pb collision at $\sqrt{s_{_{\mathrm{NN}}}}=2.76$ TeV. The shape of the $v_n$ distributions is consistent with Gaussian fluctuation in central collisions (0-2\% centrality range) for $v_2$ and over the full centrality range for $v_3$ and $v_4$. The ratio of the RMS to the mean, $\sigma_{v_n}/\langle v_n\rangle$, is studied as a function of $\langle N_{\mathrm{part}}\rangle$ and $\pT$. The values of $\sigma_{v_n}/\langle v_n\rangle$ are found to be independent of $\pT$, suggesting that the hydrodynamic response to the eccentricity of the initial geometry has little $\pT$ dependence, however they are found to reach a minimum of 0.34 for $v_2$ around $\langle N_{\mathrm{part}}\rangle\sim200$. A comparison of the $v_n$ distributions with the eccentricity distributions of the initial geometry from the Glauber and MC-KLN models, shows that both models fail to describe the data across the full centrality range. The $v_n$ from the event plane method is found to lie in between $\langle v_n\rangle$ and $\sqrt{\langle v_n\rangle^2+\sigma_{v_n}^2}$. These results may shed light on the nature of the fluctuations of the created matter in the initial state as well as the subsequent hydrodynamic evolution.

This work is in part supported by NSF under award number PHY-1019387.
%\section*{References}

\end{document}